\documentclass[twocolumn,pra,showpacs,groupedaddress,assymb,amsmath]{revtex4}
\usepackage{}
\usepackage{graphicx}
\usepackage{amssymb}
\usepackage{amsmath}
\newcommand{\Tr}{\text{Tr}}
\usepackage{extarrows}
\begin{document}
\title{Quantum response theory for open systems and its application to Hall conductance}
\author{H. Z. Shen$^{1,2}$, M. Qin$^{1,2}$, Y. H. Zhou$^{1,2}$,
X. Q. Shao$^{1}$, and X. X. Yi$^1$}\email{yixx@nenu.edu.cn} \affiliation{$^1$Center for Quantum Sciences and
School of Physics, Northeast
Normal University, Changchun 130024, China\\
$^2$ School of Physics and Optoelectronic Technology\\
Dalian University of Technology, Dalian 116024 China}

\date{\today}

\begin{abstract}
Quantum linear  response theory
considers only the response of a closed quantum system to a
perturbation up to   first order in the perturbation. This theory
breaks down when the system subjects to environments and the
response up to  second order in perturbation is not negligible. In
this paper, we develop a quantum nonlinear response theory for open
systems. We first formulate this theory in terms of general
susceptibility, then apply it to deriving the Hall conductance for
the open system at finite temperature. Taking the two-band model as
an example, we derive the Hall conductance for the two-band model.
We calculate the Hall conductance for a two-dimensional
ferromagnetic electron gas and a two-dimensional lattice model via
different expressions for $d_{\alpha}(\vec p), \ \alpha=x,y,z$. The
results show that the transition points of topological phase almost
remain unchanged in the presence of environments.
\end{abstract}

\pacs{73.43.Cd, 03.65.Yz, 03.65.Vf, 73.20.At} \maketitle
\maketitle
%73.43.Cd: Theory and modeling
%03.65.Yz: Decoherence; open systems; quantum statistical methods (see also 03.67.Pp in quantum information; for decoherence in Bose-Einstein condensates, see 03.75.Gg)
%03.65.Vf: Phases: geometric; dynamic or topological
%73.20.At: Surface states, band structure, electron density of states
\section{introduction}
Applying a small perturbation to a quantum system, we may compute the response of
some operators of the system  to the perturbation, for example the
electric current density as a response to an electric field. When
consider  the response only to first order in the perturbation, the
theory is the so called quantum linear response theory.

The Kubo Formula is an equation which expresses the linear response
of an observable quantity due to a time-dependent perturbation,
which  has been widely used in condensed matter physics
\cite{thouless82,Kohmoto1985343, qi2006085308,zhou2006165303} since
it was first derived by Kubo in 1957 \cite{kubo1957570}.  However,
Kubo's theory is only  valid for an equilibrium closed system in the
linear regime\cite{note1}. In recent years, linear response theory
for open systems has attracted more and more attentions in
biophysics, nano-physics and condensed matter physics. A linear
response theory based on the master equation
 \cite{Avron19996097,Avron2011053042,Avron2012800,
Uchiyama2009021128,Saeki0311312010,Kolovsky201150002} and the
hierarchical equation of motion \cite{Jin2008234703,Wei20115955} has
been developed for open systems. The key issue and difference of
those approaches are how to get the reduced density matrix of the
open system---while the first approach obtains the density matrix by
master equations, the second by the Hierarchical equation.
Both approaches need to calculate
the density matrix or the dynamics of the density matrix. This means
we have to trace out the environment first, then calculate the
response---a treatment on nonequal footing for the system and
environment. It is worth addressing that, based on  the reduced
density matrix,  a response theory for the open system was
developed, we refer the readers to \cite{shen20146455} for details.

Treating the system and the environment on equal footing, we here
develop a quantum response theory for  open system by a different
approach. This approach  does not need to get the reduced density
matrix and the theory is beyond the linear response  regime
 \cite{Uchiyama2009021128,Saeki0311312010}.
Our treatment is not limited to a
specific  system. Rather, it is applicable to any strength  of
perturbation and  open quantum systems. Hence it is suitable for
strong external fields applied to topological
insulators \cite{Peterson39691967}. We will apply this theory to
topological insulators (TIs) described by the two-band model and
study the effect of environment on the Hall conductance at both zero
and finite temperatures.

Topological insulators are a broad class of unconventional materials
that are insulating in the interior but conduct along the
edges \cite{konig07766000, Bernevig06,hsieh08970, roth09294}. Over
the last decades, topological insulators  have attracted a great
deal of interest due to their interesting  features and possible
application in quantum computation. Recently, efforts have also been
made to investigate topological insulator for open systems, e.g.,
density-matrix Chern insulators by thermal
noise \cite{rivas13,viyuela2012}, zero-temperature Hall conductance
subjected to decoherence \cite{shen20146455}, topological order by
dissipation \cite{Bardyn135135}. These stimulate  us to develop a
response theory for open systems to high order in perturbation, and
apply it to study topological insulators subjected to environments.

The remainder of the paper is organized as follows. In Sec. {\rm
II}, we extend the linear response theory of  closed  system to open systems and
nonlinear case. In Sec. {\rm III}, as an application of our theory, we derive the finite-temperature Hall conductance
for open systems and exemplify it into the two-band model. The Hall
conductance for a two-dimensional ferromagnetic electron gas and a
two-dimensional lattice model is also discussed in this section. Discussion and conclusions are given in Sec. {\rm IV}.
\section{Nonlinear quantum response theory for quantum open systems}
Consider a  quantum system described by a time-dependent Hamiltonian
$H_S(t)$ coupling to   environment  $H_R$ in an external  field
$H_e(t)$, the total  Hamiltonian ${H_T}(t)$ reads,
\begin{equation}
{H_T}(t) = {H_S(t)} + {H_R} + {H_{SR}} + \varepsilon {H_{e}}(t),
\label{HTt}
\end{equation}
where $H_{SR}$ denotes  the interaction  between the system and the
environment. $\varepsilon $ stands for the field  strength, and
${H_{e}}(t) = - \sum\limits_\nu {{f_\nu }(t){C_\nu }}\label{Hex}$.
To simplify the notations in later discussions, we set
$H(t)={H_0(t)} + {H_{SR}}$, $H_0(t)={H_S(t)} + {H_R}$, and use the
notation, $L_X(t)\rho (t) = [H_X(t),\rho (t)]$ with $X=T, e,0$, or
left blank. We are interested in  the response of the open system to
the external fields $f_\nu$, which will be derived in the following.

The total density matrix $\rho _T(t)$ satisfies the Liouville
equation ${{\dot \rho }_T}(t) =  - \frac{i}{\hbar }{L_T}(t){\rho
_T}(t) =  - \frac{i}{\hbar }L(t){\rho _T}(t) - \frac{i}{\hbar
}\varepsilon {L_e(t)}{\rho _T}(t)$. By dividing the total density
matrix into two parts ${\rho}(t)$ and $\rho _{Te}(t)$ with
${\rho}(t)$ satisfying ${{\dot \rho }}(t) =  - \frac{i}{\hbar
}{L}(t){\rho}(t)$, we have
\begin{eqnarray}
{\rho _e}(t) =  - \frac{i}{\hbar }\varepsilon  \Tr_R\int_{{t_0}}^t
{g(t,u)} {L_e}(u){\rho _T}(u)du,\label{rhoet}
\end{eqnarray}
where ${\rho _{e}}(t)$ was defined as ${\rho
_{e}}(t)=\Tr_R\left(\rho _{Te}(t)\right )$ and $\rho _{Te}(t)$ is
defined as a change  of the density matrix due to the external
field. $\Tr_X$ denotes a trace over $X$.

Eq.~(\ref{rhoet}) completely describes   the influence of the
external fields on the  system subjected to an environment. $g(t,u)$
is the chronological time-ordering operator, $g(t,u) = T_+\exp
\left[ { - \frac{i}{\hbar }\int_u^t {dsL(s)} } \right]$, $(t > u)$. Now
we can define the general nonlinear response tensor ${\chi _{\mu \nu
}}(t,u)$ for the open system via the change of the expectation value
of a system operator $F_\mu$ caused by the external  field,
$\left\langle {{F_\mu }(t)} \right\rangle
_{e}=\Tr_S(\rho_e(t)F_\mu)$,
\begin{eqnarray}
{\left\langle {{F_\mu }(t)} \right\rangle _{e}} \equiv
\sum\limits_\nu {\int_{t_0}^t{du{\chi_{\mu \nu }}(t,u){f_\nu }(u)}
}, \label{AJt}
\end{eqnarray}
where the  nonlinear response tensor $\chi _{\mu \nu }(t,u)$ is
given by
\begin{eqnarray}
\chi _{\mu \nu }(t,u) = \frac{i}{\hbar }\varepsilon  \Tr_{SR}\{
{F_\mu }g(t,u)[{C_\nu },x(u)]\}, \label{chits}
\end{eqnarray}
with
\begin{eqnarray}
x(u) &=& \rho (u) + \sum\limits_{n = 1}^\infty  {( - } \varepsilon {)^n}\int_{{t_0}}^u {d{t_1}} \int_{{t_0}}^{{t_{n - 1}}} {d{t_n}} \nonumber\\
&& \cdot s(u,{t_1}) \cdots s({t_{n - 1}},{t_n})\rho ({t_n}),
\label{volterraexpansion}
\end{eqnarray}
and $s (t,u) = \frac{i}{\hbar }g(t,u){L_{e}}(u)$.

In the following, we will restrict ourself to consider a
time-independent system. The formalism can be easily generalized to
time-dependent systems.  Suppose the whole system is in an
equilibrium state $\rho_{eq}$ at temperature $T$. The main task of
the general nonlinear response theory is  to calculate the
susceptibility,
\begin{eqnarray}
{\chi _{\mu \nu }}(\omega ) &=& \int_0^\infty {dt{e^{i\omega t}}{\chi
_{\mu \nu }}(0, - t)} \nonumber\\
&=& \frac{i}{\hbar }\varepsilon \int_0^\infty
{dt{e^{i\omega t}}\Tr_{SR}} \{ {F_\mu }{e^{ - \frac{i}{\hbar
}Lt}}[{C_\nu },x( - t)]\}\nonumber\\
\label{chiomege}
\end{eqnarray}
with an assumption that $H_{e}$ is turned on at ${t_0} \to - \infty$
and the steady response outputs at $t=0$. Defining ${q_\nu }(t) =
\Tr_R{e^{ - \frac{i}{\hbar }Lt}}[{C_\nu },x( - t)]$, we rewrite the
susceptibility as,
\begin{equation}
{\chi _{\mu \nu }}(\omega ) = \frac{i}{\hbar }\varepsilon
\int_0^\infty  {dt{e^{i\omega t}}\Tr_S} [{F_\mu }{q_\nu }(t)].
\end{equation}
By the modified Laplace transformation  $g(\omega ) = \int_0^\infty
{dt{e^{ i\omega t}}g(t)}$ \cite{zhang2012170402}, we find,
\begin{eqnarray}
{\chi _{\mu \nu }}(\omega ) &=&  \varepsilon \chi _{\mu \nu
}^{(1)}(\omega ) + \sum\limits_{n = 2}^\infty  {{\varepsilon ^n}\chi
_{\mu \nu }^{(n)}(\omega )} , \label{appropria}
\end{eqnarray}
where $\chi _{\mu \nu }^{(n)}(\omega ) =  \frac{i}{\hbar
}\Tr_S[{F_\mu }q_\nu ^{(n)}(\omega )]$ with
\begin{eqnarray}
q_\nu ^{(n)}(\omega ) &=& \int_0^\infty  {dt} \exp (  i\omega
t)q_\nu
^{(n)}(t),\nonumber\\
q_\nu ^{(1)}(t) &=& \Tr_R{e^{ - \frac{i}{\hbar }Lt}}[{C_\nu },{\rho
_{eq}}],
\end{eqnarray}
 and
\begin{small}
\begin{eqnarray}
q_\nu ^{(n)}(t) &=& {( - 1)^{n + 1}}\Tr_R\{ {e^{ - \frac{i}{\hbar
}Lt}}[{C_\nu },\int_{ - \infty }^{ - t} {d{t_1}}
\cdot\cdot\cdot\int_{{-\infty}}^{{t_{n - 2}}} {d{t_n}}\nonumber\\
&& \cdot s( - t,{t_1})
\cdots s({t_{n - 2}},{t_{n - 1}}){\rho _{eq}}]\} ,(n \ge
2).\label{qnu}
\end{eqnarray}
\end{small}

Here the first term in Eq.~(\ref{appropria}) represents  the linear
response, while the others are  nonlinear responses  of the open
system. Eq.~(\ref{appropria}) suggests  that in order to get the
nonlinear susceptibility, we have  to calculate  $q_{\nu}(\omega )$
through $q_{\nu}(t)$.  The time evolution of $q_{\nu}(t)$ is given
by (see Appendix B),
\begin{equation}
{{\dot q_{\nu}}}(t) =  - \frac{i}{\hbar }[{H_S},{q_{\nu}}(t)] +
\int_{t_0}^t {c(t - \tau ){q_{\nu}}(\tau )d\tau  + K(t)},
\label{nonexactproject}
\end{equation}
where  the kernel  $c(t)$ and  term $K(t)$ are given by $c(t) = -
\Tr_R[L{e^{ - \frac{i}{\hbar }QLt}}QL{\rho_R}/{\hbar ^2}]$ and
\begin{eqnarray}
K(t) &=&  - \frac{i}{\hbar }\Tr_R[L{e^{ - iQLt/\hbar }}Q\Lambda
(0)]+\Tr_R{e^{ - \frac{i}{\hbar }Lt}}\dot \Lambda (t) \nonumber\\
&&- \frac{i}{\hbar }\Tr_R[L\int_{{0}}^t d \tau {e^{ - \frac{i}{\hbar
}QL(t - \tau )}}Q{e^{ - \frac{i}{\hbar }L\tau }}\dot \Lambda (\tau
)],\nonumber\\
\label{ktt}
\end{eqnarray}
respectively.   $\Lambda (t)= [{C_\nu},x( - t)]$, and $x(t)$ is
defined  by Eq.~(\ref{volterraexpansion}).

In Appendix C, we use an example---a
single-mode cavity system coupled to environment in a time-dependent
external field---to illustrate the  nonlinear response when the
external field is not weak. In addition,  we show that it is
difficult to derive analytically the nonlinear response of a general
open system to a perturbation at finite temperatures. Considering
the fact that  two-dimensional topological insulators described by
the two-band model has been experimentally observed
 \cite{konig07766000, Bernevig06,hsieh08970, roth09294}, we will
focus on the linear response theory for an  open system in the
following.

We assume the system and environment
be initially in their thermal equilibrium,  $\rho(0)={\rho_{eq}} =
{e^{ - \beta H}}/\Tr_{SR}{e^{ - \beta H}}$. To calculate the  linear
response,    we take the first term in Eq.~(\ref{volterraexpansion})
as $x(u) \equiv {\rho _{eq}}$. Furthermore, we take  ${\rho _{eq}} =
{\rho _S} \otimes {\rho _R} + O({H_{SR}})$ for simplicity,  the term
$K(t)$ in Eq.~(\ref{nonexactproject}) vanishes since
$Q[{C_\nu},{\rho _{eq}}] = Q[{C_\nu},{\rho _S}]{\rho _R} = 0$. After
the modified Laplace transformation for Eq.~(\ref{nonexactproject})
and expanding it to  second order in the system-environment
couplings $H_{SR}$, we have
\begin{eqnarray}
- \frac{i}{\hbar }{L_S}{q_\nu }(\omega ) +  [i\omega  + c(\omega
)]{q_\nu }(\omega ) =  - {q_\nu }(0),\label{reduceslaplace}
\end{eqnarray}
where
\begin{equation}
\begin{aligned}
c(\omega ) = \frac{{ - 1}}{{{\hbar ^2}}} \int_0^\infty  {dt}
{\left\langle {{L_{SR}}{\cal H}_{SR}^I(t)} \right\rangle _R}\exp
[i(\omega - {L_S}/\hbar )t],\label{reducesto2}
\end{aligned}
\end{equation}
with  ${\cal H}_{SR}^I(t) = \exp [-i{L_0}t/\hbar ]{H_{SR}}$. Here
and hereafter, the perturbation parameter  $\varepsilon$ is absorbed
in $C_\nu$. Eq. (\ref{reduceslaplace}) and Eq. (\ref{appropria}) is
one of the main results in this paper. Eq. (\ref{appropria}) is the
response of the open system to the external field, and Eq.
(\ref{reduceslaplace}) is the key element to calculate the linear
response.

\section{Application to  Hall conductance in the two-band model.}
We now apply the quantum response theory for open systems to a
two-band model. The model Hamiltonian is,
\begin{eqnarray}
{H_S} = \sum\limits_{\vec p} {H_S(\vec p)} ,H_S(\vec p) =
\varepsilon (\vec p) + \sum\limits_{\alpha  = x,y,z} {{d_\alpha
(\vec p)}{\sigma _\alpha }}, \label{fintehp}
\end{eqnarray}
where $\varepsilon (\vec p)=p^2/2m^*$ denotes  the kinetic energy
with the band electron effective mass $m^*$ and ${{\sigma _\alpha
}}$ are the Pauli matrices. $\vec p = ({p_x},{p_y})$ stands for the
Bloch wave vector of the electron.

The two-band system is an
idealization but can be realized approximatively with ultra-cold
atoms \cite{Bloch802008} using different techniques as, e.g.,
super-lattices \cite{Salger1904052007,Salger3262009,Breid81102006},
with hole-related band  suggested by Raghu et al. in
 \cite{Raghu2205032008} and in spin-$1/2$ electrons   with  the
spin-orbit coupling \cite{qi2006085308}. Besides the possibility of
experimental realization, the two-band model is also interesting as
a simple model in condensed matter physics.

The eigenenergies of the two-band Hamiltonian are, ${E_m}(\vec p) =
\varepsilon (\vec p) + md(\vec p)$, with $m = \pm $ and $d = \sqrt
{d_x^2 + d_y^2 + d_z^2}$. The corresponding eigenstates take
\begin{eqnarray}
\left| { + (\vec p)} \right\rangle  = \left( {\begin{array}{*{20}{l}}
{\cos \frac{\theta }{2}{e^{ - i\phi }}}\\
{\begin{array}{*{20}{c}}
{}
\end{array}\sin \frac{\theta }{2}}
\end{array}} \right),\left| { - (\vec p)} \right\rangle  = \left( {\begin{array}{*{20}{l}}
{ - \sin \frac{\theta }{2}{e^{ - i\phi }}}\\
{\begin{array}{*{20}{c}}
{}
\end{array}\begin{array}{*{20}{c}}
{}
\end{array}\cos \frac{\theta }{2}}
\end{array}} \right),\label{finiteeigenstate}
\end{eqnarray}
where $\cos \theta = {d_z}/d$, $\tan \phi = {d_y}/{d_x}$ were
defined.

The Hall conductivity tensor ${\sigma _{\mu \nu }}$  can be
calculated through  the current density ${J_\mu }(t)$ in the
$\mu$-direction ($\mu=x,y,z$),  as a response to the external
electric field ${E_\nu }(t) = {\rm{Re}}[E_\nu\exp (i\omega t)]$ in
that direction, the current density reads,
\begin{eqnarray}
{J_\mu }(t) = \mathop {\lim }\limits_{\omega  \to 0} {\rm{Re}}
[{\sigma _{\mu \nu }}(\omega ){E_\nu }\exp ( i\omega
t)].\label{finiteJmu}
\end{eqnarray}
Consider a  total Hamiltonian in the momentum space \cite{rivas13}
\begin{equation}
\begin{aligned}
{H_T}(t) = {H_S} + {H_R} + {H_{SR}} + {H_e}(t),
\label{finitehrhe}
\end{aligned}
\end{equation}
with ${H_R} = \sum\limits_j {\hbar {\omega _j} b_j^\dag {b_j}}
,{H_{SR}} = \sum\limits_j {\hbar {g_j}\sigma_- b_j^\dag }  + H.c.,$
and ${H_e}(t) = - \sum\limits_\nu  {{{\rm{P}}_\nu }} {E_\nu }(t)$.
Here ${H_S}$ is given by Eq.~(\ref{fintehp}). Eq.~(\ref{finitehrhe})
denotes the Hamiltonian of a system composed  of many particles with
charges $q_j$ and position operators $r^j={i\hbar \frac{\partial
}{{\partial p ^j}}}$ subjected to environmental noises $H_R$,
${{\rm{P}}_\nu }$ is the polarization $ {{\rm{P}}_\nu } =
\sum\limits_j {i\hbar {q_j}\frac{\partial }{{\partial p_\nu^j}}} $.
In the linear response theory, the Hall conductance
takes \cite{Stadelmann2006PHD}
\begin{eqnarray}
\begin{aligned}
{\sigma _{\mu \nu }}(\omega ) =& \frac{1}{{i\hbar }} \int_0^\infty  {dt}
\left\langle {[{{\rm{P}}_\nu }(0),{J_\mu }(t)]} \right\rangle {e^{  i\omega t}} \\
=& \frac{-V}{{\hbar \omega }}\int_0^\infty  {dt}  \left\langle
{[{J_\nu }(0),{J_\mu }(t)]} \right\rangle {e^{  i\omega t}},
\label{finitesigmamv1}
\end{aligned}
\end{eqnarray}
where $J_\nu = \dot {\rm P}_\nu /V$ and $V$ denotes the mode volume
for system. Using the relations $J=ev/V$ and ${v_\mu }(t) =
{e^{iHt}}v_\mu{e^{{\rm{ - }}iHt}}$,  we  obtain,
\begin{eqnarray}
\begin{aligned}
{\sigma _{\mu \nu }}(\omega ) =& \frac{{{-e^2}}}
{{\hbar \omega V}}\int_0^\infty  {dt} \left\langle {[{v_\nu }(0),{v_\mu }(t)]}
\right\rangle {e^{ i\omega t}}\\
=& \frac{{ {e^2}}}{{\hbar \omega V}}\int_0^\infty  {dt}
T{r_{S+R}}\{ {v_\mu }{e^{ - \frac{i}{\hbar }Lt}}[{v_\nu },{\rho _S}]\} {e^{  i\omega t}} \\
&+{\cal O}({H_{SR}})\\
\simeq & \frac{{ {e^2}}}{{\hbar \omega V}}T{r_S}[{v_\mu }{q_\nu
}(\omega )]. \label{finito}
\end{aligned}
\end{eqnarray}
Here we set $C_{\nu}=v_\nu$, therefore  $q_{\nu}(0)= [{v_\nu },{\rho
_S}] + {\cal O}({H_{SR}})$ .
Noticing  Eq.~(\ref{finitehrhe}) and taking the
Lorentzian spectrum density $J(\omega ) = \frac{\Gamma }{{2\pi
}}\frac{{{\lambda ^2}}}{{{{\omega }^2} + {\lambda ^2}}}$
 \cite{breuer20021} into consideration, under the Markovian approximation we obtain from
Eq.~(\ref{reducesto2})
\begin{equation}
c(\omega ){q_\nu }(\omega ) =  \Gamma [2{\sigma _ - }{q_\nu }(\omega
){\sigma _ + } - {\sigma _ + }{\sigma _ - }{q_\nu }(\omega ) -
{q_\nu }(\omega ){\sigma _ + }{\sigma _ - }].\label{reducesr2}
\end{equation}
Substituting Eq.~(\ref{reducesr2}) into Eq.~(\ref{reduceslaplace}),
and inserting the identity $\sum\limits_m {\left| {m(\vec p)}
\right\rangle \langle m(\vec p)| = I} $ into the result, we obtain
\begin{equation}
\begin{aligned}
0=&{S_{\nu ,nm}} + i{q_{\nu ,nm}}[\hbar \omega  - {E_n}(\vec p) +
{E_m}(\vec p)] \\
&+ 2\Gamma \hbar \sum\limits_{\vec p,ij} {{{({\sigma _
- })}_{ni}}} {q_{\nu ,ij}}{({\sigma _ + })_{jm}} - {F_{\nu ,nm}} -
F_{\nu ,mn}^\dag  ,\label{finitedetails}
\end{aligned}
\end{equation}
where the coefficient ${F_{\nu,nm}} =  \hbar \Gamma
\sum\limits_{{\vec p},j} {{q_{\nu,{nj}}}} {({\sigma _ + }{\sigma _ -
})_{jm}}$, ${S_{\nu,nm}} = \hbar [f_m({\vec p}) - f_n({\vec
p})]{\upsilon _{\nu ,nm}}$ with the definition ${y_{mn}} = \langle
m({\vec p})|y\left| {n({\vec p})} \right\rangle$. Here we have
applied the fact that ${\rho _S} = \sum\limits_j {\left| {j({\vec
p})} \right\rangle {f_j}} ({\vec p})\langle j({\vec p})|$, and
$f_j({\vec p}) = 1/\{ \exp [\beta ({E_j}({\vec p}) - \mu )] + 1\} $
is the Fermi-Dirac distribution function with $\mu$ the chemical
potential, $\beta=1/k_{B}T$, $k_B$ the Boltzmann constant, and $T$
the temperature. Tedious but straightforward algebra yields ($m \ne
n$),
\begin{equation}
\begin{aligned}
{q_{\nu ,nm}} = \frac{{{S_{\nu ,mn}}{A_{nm}} - \hbar \Gamma {B_{nm}}cos2\theta  - {S_{\nu ,nm}}(\hbar \omega  + {\rm{i}}\hbar \Gamma )\hbar \Gamma }}{{2[{D_{nm}} + 2{{(\hbar \omega )}^3} + 8{\rm{i}}{{(\hbar \omega )}^2}\hbar \Gamma  - {\rm{i}}\hbar \Gamma e_{nm}^2cos2\theta ]}},\label{finitealphanm}
\end{aligned}
\end{equation}
where the coefficients take ${e_{nm}} =  {E_n}({\vec p}) -
{E_m}({\vec p}),{A_{nm}} = 4{\rm{i}}{(\hbar\omega )^2} - \hbar\omega
(11\hbar\Gamma  + 4{\rm{i}}{e_{nm}}) - \hbar\Gamma
(7{\rm{i}}\hbar\Gamma  - 6{e_{nm}}),{B_{nm}} =
{S_{\nu,nm}}(\hbar\omega + {\rm{i}}\hbar\Gamma ) +
{S_{\nu,mn}}(\hbar\omega  + {\rm{i}}\hbar\Gamma  -2{e_{nm}})$, and
${D_{nm}} = -2\hbar\omega [5{(\hbar\Gamma) ^2} + e_{nm}^2] -
{\rm{i}}\hbar\Gamma [4{(\hbar\Gamma )^2} + 3e_{nm}^2]$.

The diagonal elements of $q_{\nu}(\omega)$  is not listed here,
since it has no contribution to the Hall conductivity. In the weak
dissipation limit, $\Gamma  \to 0$, we can expand $q _{\nu,nm}$ in
powers of $\Gamma$. To first order in $\Gamma$, $q _{\nu,nm}$ can be
written as,
\begin{eqnarray}
{q_{\nu ,nm}} = q_{\nu,nm}^{(0)} + \hbar \Gamma q_{\nu,nm}^{(1)},\label{finitealpha01}
\end{eqnarray}
where the zeroth- and first-order of $q_{\nu,nm}$  take,
\begin{equation}
\begin{aligned}
q_{\nu,nm}^{(0)} =& \frac{{i{S_{\nu,nm}}}}{{\hbar\omega  - {e_{nm}}}},\\
q_{\nu,nm}^{(1)} = &\frac{{g(\theta ){S_{\nu,nm}}}} {{4{{(\hbar
\omega  - {e_{nm}})}^2}}} + \frac{{h(\theta
){S_{\nu,mn}}}}{{4({\hbar ^2}{\omega ^2} - e_{nm}^2)}},
\label{finite0gamma1}
\end{aligned}
\end{equation}
respectively.  $g(\theta ) =5- \cos 2\theta ,h(\theta ) = 1-\cos
2\theta  $. Substituting Eq.~(\ref{finitealpha01}) into the third
equation of Eq.~(\ref{finito}), we have
\begin{eqnarray}
{\sigma _{\mu \nu }}(\omega ) = \sigma _{\mu \nu }^{(0)}(\omega )  +
\hbar \Gamma \sigma _{\mu \nu }^{(1)}(\omega ), \label{sigmaomega}
\end{eqnarray}
with the zeroth- and first-order Hall conductivity at
finite-temperature $T$,
\begin{equation}
\begin{aligned}
\sigma _{\mu \nu }^{(0)}(\omega )=& \frac{{{e^2}}}
{{\hbar \omega V}}\sum\limits_{{\vec p},m \ne n}
{\langle m{\rm{(}}{\vec p}{\rm{)}}|{\upsilon _\mu }\left| {n{\rm{(}}{\vec p}{\rm{)}}}
 \right\rangle } \left( {\frac{{i{S_{\nu ,nm}}}}{{\hbar \omega  - {e_{nm}}}}} \right),\\
\sigma _{\mu \nu }^{(1)}(\omega )=&   \frac{{{e^2}}}
{{\hbar \omega V}}\sum\limits_{{\vec p},m \ne n}
{\langle m{\rm{(}}{\vec p}{\rm{)}}|{\upsilon _\mu }\left| {n{\rm{(}}{\vec p}{\rm{)}}}
 \right\rangle [} \frac{{g(\theta ){S_{\nu ,nm}}}}{{4{{(\hbar \omega  - {e_{nm}})}^2}}}\\
 &+ \frac{{h(\theta ){S_{\nu ,mn}}}}{{4({\hbar ^2}{\omega ^2} - e_{nm}^2)}}].
\label{fensigma1}
\end{aligned}
\end{equation}
Finally, we can obtain the
finite-temperature Hall conductance for the open system in the weak
dissipation limit, (i.e., $\Gamma  \to 0$, see Appendix D)
\begin{eqnarray}
{\sigma _{\mu \nu }} =  \sigma _{\mu \nu }^{(0)} + \sigma _{\mu \nu
}^{(1)}\equiv C\frac{{{e^2}}}{h}, \label{totalsigma}
\end{eqnarray}
where
%\begin{small}
\begin{equation*}
\sigma _{\mu \nu }^{(0)} = \frac{{{e^2}}}{{2\hbar }}\int
\tau(\vec{k}){\varepsilon _{\alpha \beta \gamma }}
{\frac{{{d^2}k}}{{{{(2\pi )}^2}}}},
\end{equation*}
\begin{footnotesize}
\begin{equation}
 \sigma _{\mu \nu }^{(1)}
=\frac{{{\Gamma e^2}}}{{2\hbar }}\sum_{\alpha, \beta,\gamma}\int\tau
(\vec k)\left(\frac{\hbar}{4}g(\theta)\frac{d_{\alpha}
d_{\beta}}{d^2d_{\gamma}}+\frac{id}{4\omega}h(\theta)\frac{D_{\alpha\beta}}{d_\gamma}\right)
{\frac{{{d^2}k}}{{{{(2\pi )}^2}}}}\label{finitefinallyH}
\end{equation}
\end{footnotesize}
with  $\tau(\vec{k})=\frac{{[f_ + (k) - {f_ - }(k)]}}{{{d^3}}}
\frac{{\partial {d_\alpha }}}{{\partial {k_\mu }}}\frac{{\partial
{d_\beta }}}{{\partial {k_\nu}}}{d_\gamma }$, ${D_{\alpha \beta }} =
{\rm{Im}}[\langle  + (\vec p)|{\sigma _\alpha } \left| { - (\vec p)}
\right\rangle \langle  + (\vec p)|{\sigma _\beta }\left| { - (\vec
p)} \right\rangle ]$.  $C$ in Eq. ({\ref{totalsigma}) defines the Chern
number of the open system. Here we have used $\sum\limits_{\vec p}
\to \frac{V}{{{{(2\pi \hbar )}^2}}}\int {dp_xdp_y}$ and
$\vec{p}=\hbar \vec{k}.$ Note that the correction
$\sigma_{\mu\nu}^{(1)}$ in general is complex, it contains a real
part (the first term) and an imaginary part (the second term). In
the following, we will present two examples. In the first, the
correction to the Hall conductance from the system-environment
coupling is real, while it is imaginary in the second. The two
examples together exemplify the effect of the environment on the
Hall conductance.

%figure1
\begin{figure}[h]
\centering
\includegraphics[angle=0,width=0.5\textwidth]{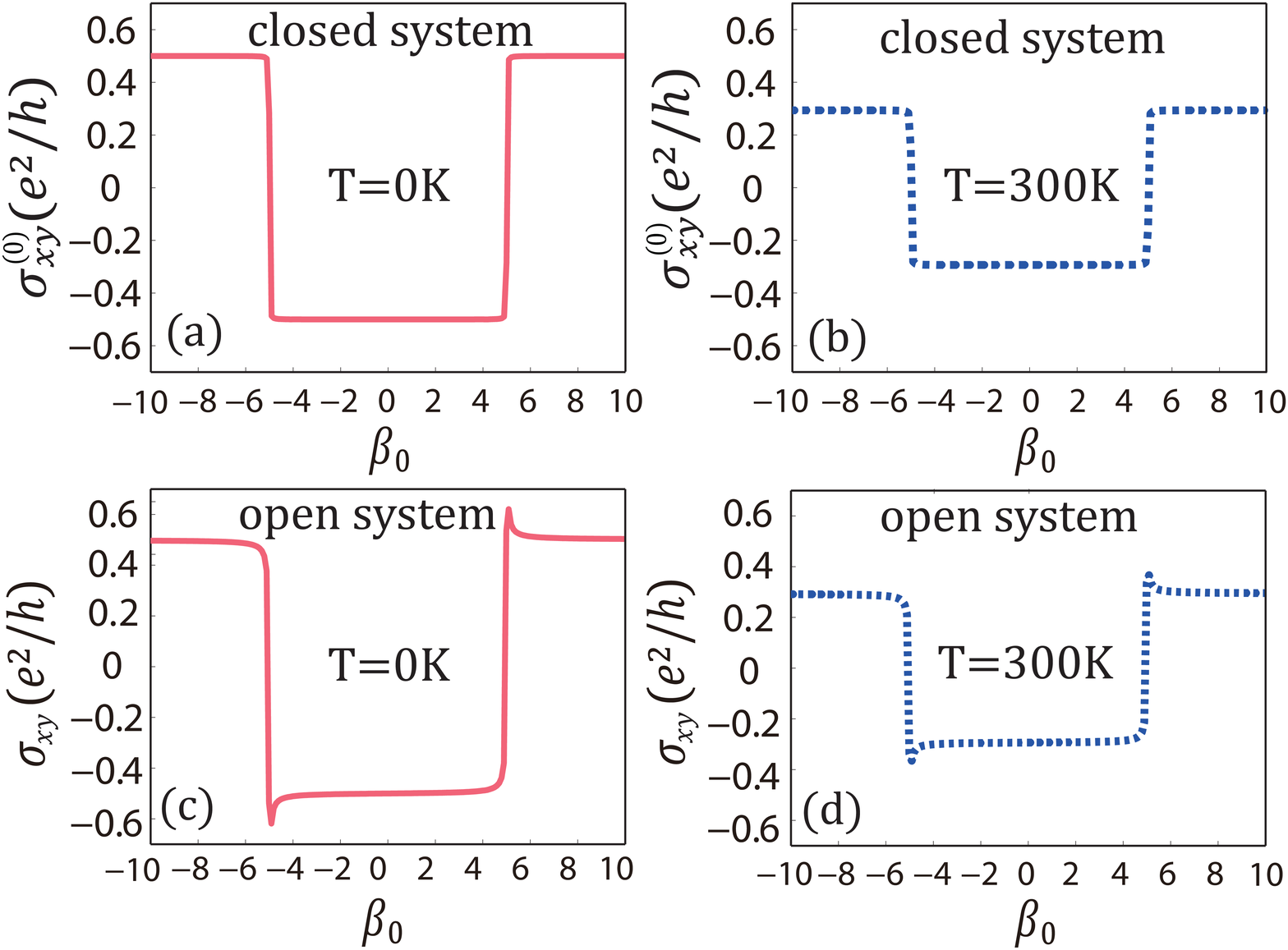}
\caption{(Color online) The zero-order conductivity $\sigma _{xy}^{(0)}$
[(a) and (b)]( i.e., $\Gamma=0$) and the total conductivity  $\sigma
_{xy}$ [(c) and (d)] as a function of $\beta_0$ (meV$ \cdot
$nm/$\hbar$). The zero-order conductivity is exactly the
conductivity of the closed system. It is worth addressing that the
imaginary part  of the Hall conductivity is
 zero in this case. Parameters chosen are $\Gamma$=0.2
meV/$\hbar$, $\lambda_0=5$ meV$\cdot $nm/$\hbar$, $\mu=1$ meV,
$h_0=2$ meV, $m^*=0.9 \text{m}_e$, $\text{m}_e$ is the mass of
electron. The temperature   $T=0 K$ for (a) and (c), and $T=300K$
for (b) and (d).} \label{rashba-D}
\end{figure}

 {\it Example 1.---} Consider a two-dimensional
ferromagnetic electron gas in the presence of both Rashba and
Dresselhaus  spin-orbit
couplings \cite{zhou2006165303,Bernevig2006236601,MEIER}. This system can
be described by Hamiltonian Eq. (\ref{fintehp}) with
$d_x=\lambda_0\hbar k_y-\beta_0\hbar k_x$, $d_y=-\lambda_0\hbar
k_x-\beta_0\hbar k_y$, and $d_z=h_0$. By the use of
Eq.~(\ref{totalsigma}), we numerically calculate the Hall
conductance and plot the   results as a function of $\beta_0$ in
Fig.~\ref{rashba-D}. Fig.~\ref{rashba-D} (a) shows a phase transition at
$\beta_0  =  \pm \lambda_0$. When ${\beta_0 ^2} > {\lambda_0 ^2}$,
the Chern number of the closed system at zero-temperature is $0.5$,
while for ${\beta_0 ^2} < {\lambda_0 ^2}$, the Chern number is
$-0.5$. This is in agreement with the analytical results of  Hall
conductivity of the closed system given by $\sigma _{xy}^{(0)}(T =
0) = \frac{1}{2}{\mathop{\rm sgn}} (\beta _0^2 - \lambda _0^2)$ in
units of $\frac{e^2}{h}$. For the open system, the plot shows that
the phase transition can still emerge. This can be found in
Fig.~\ref{rashba-D} (c),   the Chern number  at zero-temperature is
about $0.5$ when ${\beta_0 ^2}
> {\lambda_0 ^2}$, while for ${\beta_0 ^2} < {\lambda_0 ^2}$, the
Chern number is about $-0.5$. A  critical value
$\beta_0=\pm\lambda_0$ at which the Hall conductance changes
abruptly can be found. This suggests that the topological phase
transition  survives  for open system. We also find that at the
critical point $\lambda_0$, $\sigma_{xy}$ increases compared to
$\sigma_{xy}^{(0)}$, while it decreases at $-\lambda_0$. Except the
two critical points. The effect of the environment on the
conductance is almost zero.  Thermal fluctuation diminishes the
difference of the Hall conductivity between different phases, but it
does not change the nature of the phase, see Fig.~\ref{rashba-D} (b)
and (d). It is worth addressing that the imaginary part of
$\sigma_{xy}^{(1)}$ vanishes for in this model, this is different
from the example below.

%figure2
\begin{figure}[h]
\centering
\includegraphics[angle=0,width=0.5\textwidth]{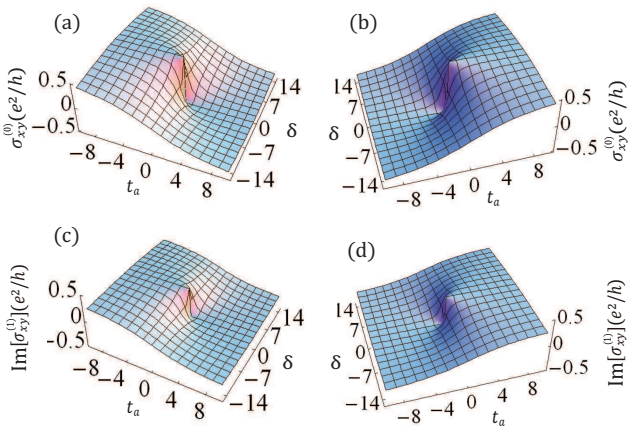}
\caption{(Color online) The zero-order  Hall conductivity $\sigma
_{xy}^{(0)}$ [(a) and (b)] and the imaginary part ${\rm{Im}}[\sigma
_{\mu \nu }^{(1)}]$ [(c) and (d)]  of the first-order Hall
conductivity as a function of $t_a$ (meV) and $\delta$ (meV) at zero
temperature. Note that the real part ${\rm{Re[}}\sigma _{\mu \nu
}^{(1)}]$ of the Hall conductivity of the open system is   zero in
this case. Define two coefficients, $m_1=4n_1+1$ or $4n_1+4$
($n_1$=0,1,2$,\cdot \cdot \cdot)$, and $m_2=4n_2+2$ or $4n_2+3$
($n_2$=0,1,2$,\cdot \cdot \cdot)$. Parameters chosen are
$\Gamma=0.1$ mev$/\hbar$, $\omega=0.2$ mev$/\hbar$, $p = 1, q= 4, l
= 1$. $m_0=m_1$, for (a) and (c). $m_0=m_2$, for (b) and (d).} \label{tight_bind}
\end{figure}
%figure3
\begin{figure}[h]
\centering
\includegraphics[angle=0,width=0.5\textwidth]{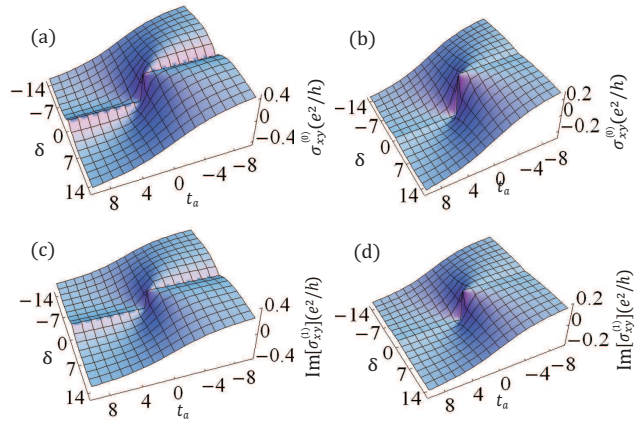}
\caption{(Color online) The  Hall  conductivity of the closed system
$\sigma _{xy}^{(0)}$ [(a) and (b)] and the imaginary part of the
first order  Hall conductivity  ${\rm{Im}}[\sigma _{\mu \nu
}^{(1)}]$ [(c) and (d)] of the open system. We plot them  as a
function of $t_a$ (meV) and $\delta$ (meV) at finite temperatures.
We address that the real part ${\rm{Re[}}\sigma _{\mu \nu }^{(1)}]$
of the Hall conductivity of the open system is  zero in this case.
Parameters chosen are $\Gamma=0.1$ mev$/\hbar$, $\omega=0.2$
mev$/\hbar$, $p = 1, q= 4, l = 1$, $m_0=m_1$, $T=30K$ for (a) and
(c).  $T=300K$ for (b) and (d).} \label{tight_bind1}
\end{figure}

{\it Example 2.---} As the second example, we consider tight-binding
electrons in a two-dimensional lattice described by the Hamiltonian
\cite{kohmoto89}\cite{note2}
\begin{equation}
H=-t_a\sum\limits_{\left\langle {i,j} \right\rangle } {_x}
c_j^\dagger {c_i}{e^{i{\theta _{ij}}}}-t_b\sum\limits_{\left\langle
{i,j} \right\rangle } {_y} c_j^\dagger {c_i}{e^{i{\theta _{ij}}}},
\label{td2}
\end{equation}
where $c_j$ is the fermion annihilation operator on  the lattice
site $j$, $t_a$ and $t_b$ denote the hopping amplitudes along the
$x-$ and $y-$ direction, respectively. Consider two branches coupled
by $|l|-$th order perturbation, the effective Hamiltonian then takes
the form  Eq.~(\ref{fintehp}) with $d_x=\delta \cos(k_yl)$,
$d_y=\delta \sin(k_y l)$, and $d_z= 2t_a
\cos(k_x+2\pi\frac{p}{q}m_0)$,  where $p$ and $q$ are integers.
$\delta$ is proportional to (is the order of) $t_b^{\left| l
\right|}.$

In Fig.~\ref{tight_bind}, we show numerically the Hall  conductance at
zero-temperature for the open system as a function of $t_a$ and
$\delta$,  the Fermi energy is set in the gap. We find that the
 Hall conductance $\sigma _{xy}^{(0)}$ and
${\rm{Im[}}\sigma _{\mu \nu }^{(1)}]$ change its sign  when  $t_a$
crosses  zero. The topological phase transition at zero temperature
survives in the open system, this can be observed by examining  the
Hall conductance, which changes from $-0.5$ and $-0.1$ to $0.5 $ and
$0.1$ when $m_0=m_1$ [in units of  $e^2/h$, see
Figs.~\ref{tight_bind} (a) and (c)]. Similar changes from $0.5$ and
$0.1$ to $-0.5$ and $-0.1$ for $m_0=m_2$ [see Figs.~\ref{tight_bind} (b)
and (d)] are found. We here address that the behavior  of the
imaginary part ${\rm{Im[}}\sigma _{\mu \nu }^{(1)}]$ of the Hall
conductivity for the open system is the same as
$\sigma_{\mu\nu}^{(0)}$ for the closed system [see
Figs.~\ref{tight_bind}-\ref{tight_bind1}] except their amplitudes. This
means that the topological properties of open
 and closed system  are same. This observation
can be explained as follows. We analytically calculate the  Hall
conductance at zero temperature and obtain,
\begin{equation}
\begin{aligned}
\sigma _{\mu \nu }^{(0)} = &A\frac{{{e^2}}} {{2h}}{\mathop{\rm sgn}}
[{\delta _{m_0{m_1}}} - {\delta _{m_0m2}}],\\
\sigma _{\mu \nu }^{(1)} = &iB\sigma _{\mu \nu }^{(0)},
\label{analy01}
\end{aligned}
\end{equation}
where $A =  - \frac{{l{t_a}}}{{\sqrt M }}, B = \frac{{\Gamma (13M +
2{\delta ^2})}}{{12\omega M}},M = 4t_a^2 + {\delta ^2} > 0.$
Especially, we find
\begin{eqnarray}
\sigma _{\mu \nu }^{(0)} &=&  - l\frac{{{e^2}}}{{2h}}{\mathop{\rm
sgn}} ({t_a}){\mathop{\rm sgn}} [{\delta _{m_0{m_1}}} - {\delta
_{m_0m2}}],\nonumber\\
\sigma _{\mu \nu }^{(1)} &=& \frac{{13i\Gamma }}{{12\omega }}\sigma
_{\mu \nu }^{(0)}
\end{eqnarray}
when $\delta=0$. This gives the phase transition points for the
closed and open system at zero temperature with $\delta=0$ [see
Fig.~\ref{tight_bind} (a) and (c), (b) and (d)]. In contrast to the system in the
last example, the correction to the conductance due to the
system-environment coupling is imaginary, which we will refer to
environment-induced reactance. This environment induced reactance
describes the energy exchange between the system and the
environment, reminiscent of the non-Markovian effect. Moreover, from
Eq.~(\ref{analy01}) we find the reactance ${\mathop{\rm
Im}\nolimits} [\sigma _{\mu \nu }^{(1)}]$ and  $\sigma _{\mu \nu
}^{(0)}$ have the same sign because $B > 0$. At finite temperatures,
for example, $T=30$K[(a) and (c)] and $300$K[(b) and (d)](see
Fig.~\ref{tight_bind1}),  the thermal effect diminishes the amplitude of the Hall
conductance compared with the value at T = 0K (Fig.~\ref{tight_bind}), but it does
not change the topological phase. %%

\section{Discussion}

We have developed a  quantum response theory for open systems beyond
the linear regime. A general nonlinear susceptibility for the open
system is derived.  This theory provides us with a formalism to
extend the notion of the finite-temperature Hall conductance form
closed to open systems.  This comes into play when studying
decoherence effects on the Hall conductivity in  open systems. We
exemplify the theory in a two-band model that describes topological
insulators, the results show that the environment affect slightly
the  topological phase transition at finite-temperature. Although
the analysis has been restricted   to two-band models, we expect
that the general response theory could be extended to higher
dimensions and nonlinear Hall conductance for many sorts  of
topological insulators.

The prediction can be observed in the six-terminal device (for
details, see Ref.\cite{Jiang2009036803}). The environment can be
simulated by the use of  B\"{u}ttiker's virtual
probes \cite{Buttiker19863020,Shi2001045123}. As  we show, the phase
transition remains in the open system. Thus, these quantum plateaus
are observable in the mesoscopic sample even with environmental
noise.

\section*{ACKNOWLEDGMENTS}
We would like to thank Prof. D. Culcer and Prof. S. Q. Shen for valuable discussions.
This work is supported by National Natural Science Foundation of
China (NSFC) under grant  Nos 11175032,  61475033, and   11204028.

\appendix
\section{The validity of the approximation $q_{\nu}(0)\simeq [{v_\nu },{\rho _S}].$}
%figure4
\begin{figure}[h]
\centering
\includegraphics[angle=0,width=0.5\textwidth]{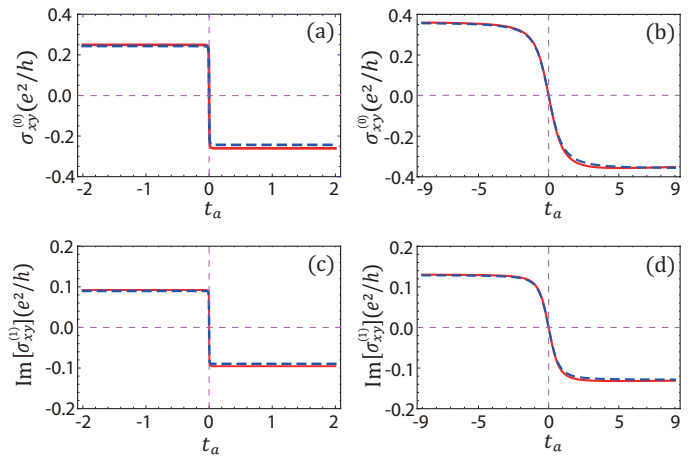}
\caption{(Color online) The Hall conductance for  closed system [(a) and
(b)] and open system [(c) and (d)] given by the third equation in
(\ref{finito}). The purpose of this figure is to show the
difference in ${\sigma _{\mu\nu}}$ caused by different order of
${q_y }(0)$ in Eq.~(\ref{appleqnu0}) by numerically solving
Eq.~(\ref{reduceslaplace}) . The red-solid and blue-dashed lines
correspond to ${q_\nu }(0) = [{v_\nu },{\rho _S}] + [{v_\nu },\rho
_S^{(2)}]$ and ${q_\nu }(0) = [{v_\nu },{\rho _S}]$, respectively.
Parameters chosen are $\omega=0.2$ mev$/\hbar$, $\lambda=25$
mev$/\hbar$, $p = 1, q= 4, L = 1$, $m_0=1$, $T=30K$, $\delta=0.01$,
$\Gamma=0$ for (a), $\delta=2$ mev, $\Gamma=0$ and (b),
$\delta=0.01$ mev, $\Gamma=0.1$ mev$/\hbar$ for (c), and $\delta=2$
mev, $\Gamma=0.1$ mev$/\hbar$ for (d).} \label{app-1}
\end{figure}

Here, we examine   the validity of the
approximating ${q_\nu }(0) = T{r_R}[{v_\nu },{\rho _{eq}}] \simeq
[{v_\nu },{\rho _S}] + O({H_{SR}})$ in Eq.~(\ref{reduceslaplace}) .
To this end, we employ the formula with two Hermitian operators $A$
and $B$,
\begin{eqnarray}
{e^{A + B}} = {e^A}{e^B}{e^{ - \frac{1}{2}[A,B]}}  \cdot
e^{(\text{high-order terms for }B)}. \label{appab}
\end{eqnarray}
Under  weak system-environment interactions $H_{SR}$,  we can expand
$\rho_{eq}$ in powers of $H_{SR}$. To second-order in $H_{SR}$ and
setting  $A=-\beta H_0$ and $B=-\beta H_{SR}$ in Eq.~(\ref{appab}),
we obtain
\begin{eqnarray}
{q_\nu }(0) = [{v_\nu },T{r_R}{\rho _{eq}}] \simeq  [{v_\nu },{\rho
_S}] + [{v_\nu },\rho _S^{(2)}], \label{appleqnu0}
\end{eqnarray}
with
\begin{eqnarray}
\rho _S^{(2)} &=& {\rho _S}[{a_1}({H_S}{\sigma _ + } {H_S}{\sigma _
- } + {\sigma _ + }{H_S}{\sigma _ - }{H_S} \nonumber\\
&&- {H_S}{\sigma _ +}{\sigma_ - }{H_S}
- {\sigma _ + }{H_S}{H_S}{\sigma _ - }) \nonumber\\
&&+ {a_2}{\sigma _ + }{H_S}{\sigma _ - }+ {a_3}{\sigma _ + }{\sigma
_ - }], \label{applerhoeq}
\end{eqnarray}
where the coefficients ${a_1} = \frac{{\Gamma \lambda  {\beta
^4}}}{{16}},{a_2} = \frac{{\Gamma \lambda {\beta ^3}}}{4}$, and
${a_3} = \frac{{\Gamma \lambda {\beta ^2}}}{4}$. Here we have used
the Lorentzian spectral density $J(\omega ) = \frac{\Gamma }{{2\pi
}}\frac{{{\lambda ^2}}}{{{{\omega }^2} + {\lambda ^2}}}$
 \cite{breuer20021}.

In order to check the validity  of  $q_\nu (0) \simeq [{v_\nu
},{\rho _S}]$ in Eq.~(\ref{reduceslaplace}), we numerically plot the
Hall conductance for closed and open system in the tight-binding
model (\ref{td2})  with different $\delta$. In Fig.~\ref{app-1}, we show a
comparison of the results with two different orders in ${q_\nu }(0)$
given in Eq.~(\ref{appleqnu0}); the simulation is performed for the
Hall conductance given by the third equation of Eq.~(\ref{finito})
with Eq.~(\ref{reduceslaplace}) . Blue-dashed line [closed system
(a) and (b) and open system (c) and (d)] in Fig.~\ref{app-1} is for ${q_\nu
}(0) = [{v_\nu },{\rho _S}]$, which are in good agreement with the
results obtained with ${q_\nu }(0) = [{v_\nu },{\rho _S}] + [{v_\nu
},\rho _S^{(2)}]$. In addition, we find the higher-order terms in
$q_\nu (0)$, which have no effects on the phase transition point and
can be ignored with respect  to the zero-order term in the two-band
model.
\section{The derivation of the equation for $\dot{q}_{\mu}(t)$.}
We apply the projection operator method to  derive
the time derivative of  $q_\nu(t)$. Note that ${q_{\nu}  }(t) =
T{r_R}{e^{ - \frac{i}{\hbar }Lt}}[{C_\nu },x( - t)]$ is not a
density matrix since the operator $C_\nu$ does not commute with $x(
- t)$ given  by Eq.~(\ref{volterraexpansion}) . To derive the equation, we
define superoperators  $P(...)=\rho_R\otimes Tr_R(...)$ and $Q=1-P$,
which are projection operators satisfying  $P^2=P$ and $Q^2=Q$. By
the standard procedure  for  deriving  a master equation  in
Ref.\cite{Uchiyama2009021128,Nakajima1958948,Zwanzig19601338}, with
two time-evolution operators defined by (the initial time is
$t_0=0$),
\begin{equation}
\begin{aligned}
W(t) = P{e^{ - \frac{i}{\hbar }Lt}}\Lambda (t),\\
Y(t) = Q{e^{ - \frac{i}{\hbar }Lt}}\Lambda (t),
\label{apprelevant}
\end{aligned}
\end{equation}
 we have
\begin{equation}
\begin{aligned}
\dot W(t) =  - \frac{i}{\hbar }PLW(t) -  \frac{i}{\hbar }PLY(t) +
P{e^{ - \frac{i}{\hbar }Lt}}\dot \Lambda (t), \label{appdiffa}
\end{aligned}
\end{equation}
and
\begin{equation}
\begin{aligned}
\dot Y(t) =  - \frac{i}{\hbar }QLY(t) -  \frac{i}{\hbar }QLW(t) +
Q{e^{ - \frac{i}{\hbar }Lt}}\dot \Lambda (t).\label{appdiffb}
\end{aligned}
\end{equation}
Solving Eq.~\ref{appdiffb}, we have
\begin{equation}
\begin{aligned}
Y(t) =& {\rm{ }}{e^{ - \frac{i}{\hbar }QLt}}Q\Lambda (0)  -
\frac{i}{\hbar }\int_0^t {{e^{ - \frac{i}{\hbar }QL(t - \tau )}}}
QLW(\tau )d\tau  \\
&+ \int_0^t {{e^{ - \frac{i}{\hbar }QL(t - \tau )}}}
Q{e^{ - \frac{i}{\hbar }L\tau }}\dot \Lambda (\tau )d\tau.
\label{appsolutionb}
\end{aligned}
\end{equation}
Substituting Eq.~(\ref{appsolutionb}) into Eq.~(\ref{appdiffa}), we obtain
\begin{equation}
\begin{aligned}
\dot W(t) = & - \frac{i}{\hbar }PLW(t) - \frac{i}{\hbar }
PL{e^{  - \frac{i}{\hbar }Q Lt}}Q\Lambda (0) - \frac{1}{{{\hbar ^2}}}
PL\\
&\cdot \int_0^t {{e^{ - \frac{i}{\hbar }QL(t - \tau )}}} QLW(\tau )d\tau - \frac{i}{\hbar }PL\\
&\cdot\int_0^t {{e^{ - \frac{i}{\hbar }QL(t - \tau
)}}} Q{e^{ - \frac{i}{\hbar }L\tau }}\dot \Lambda (\tau )d\tau  +
P{e^{ - \frac{i}{\hbar }Lt}}\dot \Lambda (t). \label{appfansoa}
\end{aligned}
\end{equation}
Finally applying   $P(...)=\rho_R \otimes Tr_R(...)$   into
Eq.~\ref{appfansoa}, we arrive at  Eq.~(\ref{nonexactproject}).
\section{Quantum response beyond the linear regime.}
As mentioned in the before,  the linear response theory is not
valid when the  external  field is not weak. Here we present an
example to show the difference between the linear response theory
and the nonlinear response theory,  it also is an illustration for
the essential role of the nonlinear response theory. We exemplify
the difference though $n_e(t)$, which is defined as the difference
in photon number with and without external fields inside  a cavity.
The results show that when the coupling between the field and system
is weak, the linear response is a good approximation, otherwise
nonlinear response should be taken into account.

Consider a  single-mode  cavity system with bare frequency
$\omega_0$ coupled to a   non-Markovian reservoir,  driven by an
external laser with frequency $\omega_L$. We assume the reservoir
modeled by a set of  harmonic oscillators is at a finite (ambient)
temperature. In a rotating frame, the Hamiltonian of the total
system  reads,
\begin{eqnarray}
{H_T} = H + {H_{e}}
\label{exactmodelht}
\end{eqnarray}
with
\begin{equation}
\begin{aligned}
H = &\hbar\Delta {a^\dag }a + \sum\limits_k {{\hbar\Omega_k}b_k^\dag {b_k}}  + \sum\limits_k
{\left( {{\hbar g_k^*}ab_k^\dag  + \hbar g_k{b_k}{a^\dag }} \right)},
\label{exactmodelh}
\end{aligned}
\end{equation}
and
\begin{equation}
\begin{aligned}
&{H_{e}}=\hbar \Omega (a + {a^\dag }),
\label{exactmodelhex}
\end{aligned}
\end{equation}
where $\Delta=\omega_0-\omega_L,$  and $\Omega_k=\omega_k-\omega_L$.
$\Omega$ is the strength of the external field, $a$ is the cavity
annihilation operator, and $b_k$ and $g_k$ are the reservoir
annihilation operator and coupling constant. In the following we
will  solve the exact non-Markovian dynamics  in Heisenberg picture.

Suppose  the system and the environment is initially
uncorrelated---the reservoir modeled by Hamiltonian ${H_R} =
\sum\nolimits_k {{\hbar\omega_k}b_k^\dag {b_k}} $ is in a thermal
equilibrium state, while the system is in a coherent state. The
initial state of the total system is,
\begin{equation}
\begin{aligned}
{\rho _{\rm{T}}}\left(0 \right) =  \left| \alpha  \right\rangle
\left\langle \alpha  \right| \otimes {\rho _R}\left( 0 \right),{\rho
_R}\left( 0 \right) = \frac{{{e^{ - \beta {H_R}}}}}{{{\rm{tr}}{e^{ -
\beta {H_R}}}}}, \label{thermal equilibrium state}
\end{aligned}
\end{equation}
where $\left| \alpha  \right\rangle$ is a  coherent state and
$\beta=1/\kappa_B T$.

Our task is to obtain  exactly the  response of the system to the
external  field and compare it with the nonlinear response. We use
the word {\it exactly} to denote that the response is not a
perturbative result---it is exact, including all orders in the
external field. With the formal solution of ${b_k}(t) =
{e^{\frac{i}{\hbar }{H_T}t }}{b_k}({0}){e^{ - \frac{i}{\hbar
}{H_T}t}}$, we rewrite the equation for $a(t)$,
\begin{equation}
\begin{aligned}
\frac{d}{{dt}}a\left( t \right) = & - i\omega_0 a\left( t \right) - \int_{0}^t {a\left(
\tau  \right)f\left( {t - \tau } \right)d\tau }  \\
&- ib(t)  - i\Omega \left(
t \right),
\label{Heisenberg equations of motion 3}
\end{aligned}
\end{equation}
where $b(t)=\sum\limits_k {{g_k}{b_k}\left( 0 \right){e^{ - i{\omega
_k}t}}}$. The memory kernel $$f\left( \tau  \right) = \sum\limits_k
{{{\left| {{g_k}} \right|}^2}{e^{ - i{\omega_k}\tau }}}\equiv \int
{d\omega J\left( \omega  \right){e^{ - i\omega \tau }}}$$
characterizes the non-Markovian dynamics of the reservoir.

Because of the  linearity of Eq.~(\ref{Heisenberg equations of
motion 3}),   $a\left( t \right)$ can be expressed as $a\left( t
\right) = u_1\left( t \right)a\left( 0 \right) + v_1\left( t
\right)$,  where  $a\left( 0 \right)$ and $b_k\left( 0 \right)$ are
the operators at the initial time. Here time-dependent coefficient
$u_1\left( t \right)$ and $v_1\left( t \right)$ can be calculated by
Eq.~(\ref{Heisenberg equations of motion 3}),
\begin{equation}
\begin{aligned}
\frac{d}{{dt}}u_1\left( t \right) =   - i\omega_0 u_1\left( t
\right) - \int_{0}^t {u_1\left( \tau  \right)f\left( {t - \tau }
\right)d\tau }, \label{coefficient equation1}
\end{aligned}
\end{equation}
\begin{equation}
\begin{aligned}
\frac{d}{{dt}}v_1\left( t \right) =& - i\omega_0 v_1\left( t \right) - \int_{0}^t {v_1\left(
\tau  \right)f\left( {t - \tau } \right)d\tau }\\
&- ib(t)  - i\Omega \left( t
\right),
\label{coefficient equation2}
\end{aligned}
\end{equation}
with  initial  conditions $u_1\left( 0 \right) = 1$ and $v_1\left( 0
\right) = 0$. $v_1\left( t \right)$  can be given analytically by
solving  the inhomogeneous equation of Eq.~(\ref{coefficient equation2}),
it leads to $v_1\left( t \right) = {\rm{ }} -
i\int_{{0}}^t {[b(\tau ) + \Omega \left( \tau \right)]u_1\left( {t -
\tau } \right)} d\tau.$

Taking the  state in Eq. (\ref{thermal equilibrium state}) into
account, we can calculate  the change of the expectation value of
the cavity photon  $n(t)=Tr_S[{a^\dag }(t)a(t)\rho(0)]$ in relevance
to the external field (\ref{exactmodelhex}) as,
\begin{eqnarray}
{n_e}(t) = {n_T}(t) - n(t) = 2{\mathop{\rm Re}\nolimits} [{u_1^ * }(t)y(t)] + {D(t)},
\label{exactnumber}
\end{eqnarray}
where $y(t) =  - i{\alpha ^ * }{D_1}(t),D(t) =  {\left| {{D_1}(t)}
\right|^2},{D_1}(t) = \Omega \int_{{0}}^t {u_1(\tau )} d\tau$.

%figure5
\begin{figure}[h]
\centering
\includegraphics[angle=0,width=0.5\textwidth]{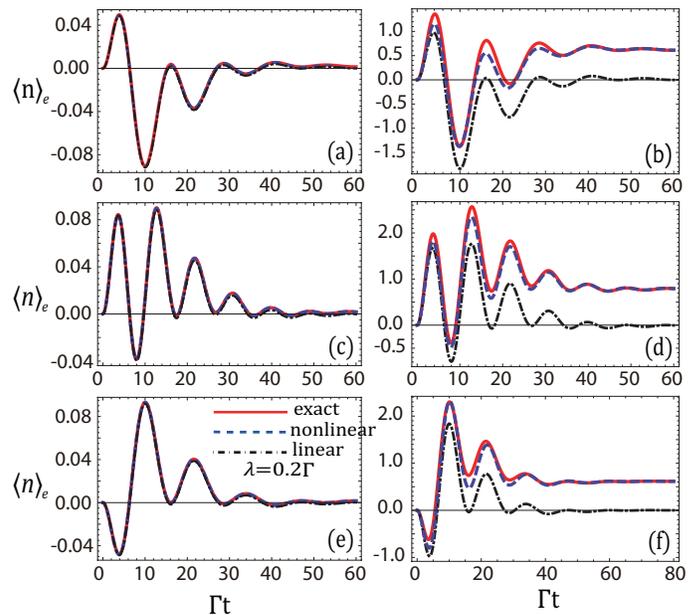}
\caption{(Color online) ${n_e}(t)$ as a function of time. The result of
linear response is given  in  (a), (c), and (e), while (b), (d), and
(f) are for two-order nonlinear response. The red solid line, blue
dashed line, and black dash-dotted line denote the exact
infinity-order response (\ref{exactnumber}), the linear response
given by Eq.~(\ref{AJt}) containing only the first term of
Eq.~(\ref{volterraexpansion}) , and the nonlinear response obtained
by Eq.~(\ref{AJt}) containing only the first two terms in
Eq.~(\ref{volterraexpansion}) , respectively. The parameters chosen
are $\lambda = 0.2\Gamma ,\Omega  = 0.01\Gamma,\alpha  =
4\Gamma,\Delta  = 0.2\Gamma$ for (a), $\Delta = 0.4\Gamma$ for (c),
$\Delta=-0.2\Gamma$ for (e). $\Omega  = 0.2\Gamma,\Delta  =
0.2\Gamma$ for (b), $\Delta  = 0.4\Gamma$ for (d),
$\Delta=-0.2\Gamma$ for (f).} \label{order1}
\end{figure}
%figure6
\begin{figure}[h]
\centering
\includegraphics[angle=0,width=0.5\textwidth]{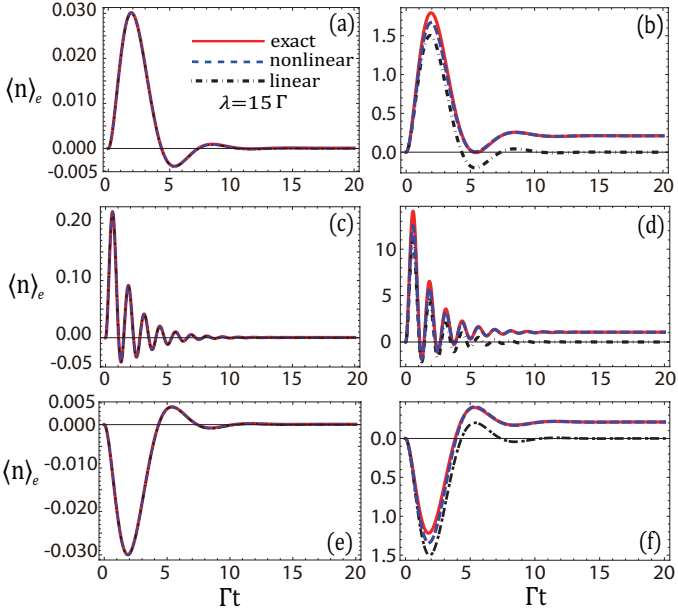}
\caption{(Color online) Comparison of the linear response  [(a), (c), and
(e)] and two-order nonlinear response [(b), (d), and (f)] in the
Markovian regime.  The red solid line, blue dashed line, and black
dash-dotted line denote the exact  response (\ref{exactnumber}), the
linear response obtained by Eq.~(\ref{AJt}) contains  only the first
term in Eq.~(\ref{volterraexpansion}), and the nonlinear response
obtained by Eq.~(\ref{AJt}) contains  only the first two terms in
Eq.~(\ref{volterraexpansion}), respectively. The parameters chosen
are $\lambda  = 15\Gamma ,\alpha  = 4\Gamma,\Delta  =
\Gamma,\Omega=0.01\Gamma$ for (a), $\Delta  =
5\Gamma$,$\Omega=0.1\Gamma$ for (c), $\Delta  =
-1\Gamma$,$\Omega=0.01\Gamma$ for (e). $\Delta  =
1\Gamma$,$\Omega=0.5\Gamma$ for (b), $\Delta  =
5\Gamma$,$\Omega=5\Gamma$ for (d), $\Delta  =
-1\Gamma$,$\Omega=0.5\Gamma$ for (f).} \label{order2}
\end{figure}

We assume that the system coupled to a reservoir  has a Lorentzian
spectral density $J(\omega ) = \frac{\Gamma }{{2\pi
}}\frac{{{\lambda ^2}}}{{{{(\omega  - {\omega _0})}^2} + {\lambda
^2}}}$ \cite{shen2013033835,zhang2013032117}. In order to examine the
validity of linear response,   we plot the response of the average
photon number to external field in three regimes divided by linear,
two-order nonlinear and exact response in Figs.~\ref{order1} and
6. In non-Markovian regime, e.g., $\lambda=0.2\Gamma$,
we can see that the results given by Eq.~(\ref{AJt})  under
the first-order approximation (containing only the first term) in
Eq~(\ref{volterraexpansion})  are in good agreement with those obtained by
the exact
 response (\ref{exactnumber}) when the interaction
strength $\Omega$  is weak [see Figs. \ref{order1} (a),
\ref{order1} (c), and \ref{order1} (e)]. With the interaction
strength $\Omega$ increasing [see Figs. \ref{order1} (b),
\ref{order1} (d), and \ref{order1} (f)], i.e., the dynamics of the
Eq. (\ref{AJt})  involving only the first and second-order
terms in Eq.~(\ref{volterraexpansion}) are in good agreement with those obtained by the
exact response  (\ref{exactnumber}), but the results obtained by the
first-order approximation (linear regime) have serious deviations
from the exact one (\ref{exactnumber}). This difference comes from
the nonlinear terms, which are ignored in linear response theory.

Examining the Markovian regime, e.g., $\lambda=15\Gamma$,  we find
that the results given by the linear response (containing only the
first term in Eq.~(\ref{volterraexpansion})  ) are in good agreement with
those obtained by the exact response (\ref{exactnumber}) when the
the interaction  $\Omega$ is weak [see
Figs. \ref{order2} (a), \ref{order2} (c), and \ref{order2} (e)].
When the the interaction strength $\Omega$ becomes strong [see
Figs. \ref{order2} (b), \ref{order2} (d), and \ref{order2} (f)],
the dynamics given by   Eq. (\ref{AJt})  involving only the first and
second-order terms in Eq.~(\ref{volterraexpansion})  are in good agreement
with those obtained by the exact expression (\ref{exactnumber}).
However, the results obtained by the first-order approximation
(linear regime) have serious deviations from those obtained by the
exact expression (\ref{exactnumber}). The same observation can be
found in the non-Markovian regime.
\section{The derivation of Eq. (\ref{totalsigma}).}
We first calculate the zeroth-order Hall conductivity at
finite-temperature by rewriting  Eq.~(\ref{fensigma1}) as
\begin{equation}
\begin{aligned}
\sigma _{\mu \nu }^{(0)}(\omega) =& \frac{{i{e^2}}}{{\omega V}}
\sum\limits_{{\vec p},m \ne n} {f_n({\vec p})[\frac{{\langle n{\rm{(}}{\vec p}{\rm{)}}|{v_\mu }
\left| {m{\rm{(}}{\vec p}{\rm{)}}} \right\rangle \langle m{\rm{(}}{\vec p}{\rm{)}}|{v_\nu }
\left| {n{\rm{(}}{\vec p}{\rm{)}}} \right\rangle }}{{\hbar \omega  + {e_{nm}}}}}\\
&- \frac{{\langle m{\rm{(}}{\vec p}{\rm{)}}|{v_\mu }\left| {n{\rm{(}}{\vec p}{\rm{)}}}
\right\rangle \langle n{\rm{(}}{\vec p}{\rm{)}}|{v_\nu }\left| {m{\rm{(}}{\vec p}{\rm{)}}}
\right\rangle }}{{\hbar \omega  - {e_{nm}}}}].
\label{finitesigma1}
\end{aligned}
\end{equation}
In the limit $\left| {\omega /{e_{nm}}} \right| \ll 1$,
\begin{eqnarray}
\frac{1}{{\hbar \omega  \pm {e_{nm}}}} = \frac{1}{{{e_{nm}}}}
\left( { \pm 1 - \frac{\hbar\omega }{{{e_{nm}}}}} \right) +\mathcal {O}(\omega^2).
\label{finiteomegaappro}
\end{eqnarray}
Substituting Eq.~(\ref{finiteomegaappro}) into
Eq.~(\ref{finitesigma1}), we have $\sigma _{\mu \nu }^{(0)}(\omega )
= {\sigma _1}(\omega) + {\sigma _2}(\omega)$ with
\begin{equation}
\begin{aligned}
{\sigma _1}(\omega ) =& \frac{{i{e^2}}}{{\omega V}}\sum
\limits_{{\vec p},m \ne n} {f_n({\vec p})} [\langle n{\rm{(}}{\vec p}{\rm{)}}|{\upsilon _\mu }\left|
{m{\rm{(}}{\vec p}{\rm{)}}} \right\rangle \langle m{\rm{(}}{\vec p}{\rm{)}}|{\upsilon _\nu }\left|
{n{\rm{(}}{\vec p}{\rm{)}}} \right\rangle  \\
&+ \langle n{\rm{(}}{\vec p}{\rm{)}}|{\upsilon _\nu }\left|
{m{\rm{(}}{\vec p}{\rm{)}}} \right\rangle \langle m{\rm{(}}{\vec p}{\rm{)}}|{\upsilon _\mu }\left|
{n{\rm{(}}{\vec p}{\rm{)}}} \right\rangle ]/{e_{nm}},\label{finitesigma11}
\end{aligned}
\end{equation}
and
\begin{equation}
\begin{aligned}
{\sigma _2}(\omega ) =& \frac{{ - i\hbar {e^2}}}{V}\sum\limits_{{\vec p},m \ne n}
{f_n({\vec p})} [\langle n{\rm{(}}{\vec p}{\rm{)}}|{\upsilon _\mu }\left|
{m{\rm{(}}{\vec p}{\rm{)}}} \right\rangle \langle m{\rm{(}}{\vec p}{\rm{)}}|
{\upsilon _\nu }\left| {n{\rm{(}}{\vec p}{\rm{)}}} \right\rangle \\
 &-
\langle n{\rm{(}}{\vec p}{\rm{)}}|{\upsilon _\nu }\left| {m{\rm{(}}{\vec p}{\rm{)}}}
\right\rangle \langle m{\rm{(}}{\vec p}{\rm{)}}|{\upsilon _\mu }\left| {n{\rm{(}}{\vec p}{\rm{)}}}
 \right\rangle ]/e_{nm}^2.\label{finitesigma12}
\end{aligned}
\end{equation}

Now we show that the first term ${\sigma _1}(\omega )$ vanishes. Set
$\mu=x$ and $\nu=y$),  we have  ${v _x} = \frac{i}{\hbar
}[{H_S},x]$, then
\begin{equation}
\begin{aligned}
\langle n{\rm{(}}{\vec p}{\rm{)}}|{v_x}\left| {m{\rm{(}}{\vec p}{\rm{)}}}
\right\rangle  =& \frac{i}{\hbar }\langle n{\rm{(}}{\vec p}{\rm{)}}|[H,x]
\left| {m{\rm{(}}{\vec p}{\rm{)}}} \right\rangle  \\
=& \frac{i}{\hbar }{e_{nm}}
\langle n{\rm{(}}{\vec p}{\rm{)}}|x\left| {m{\rm{(}}{\vec p}{\rm{)}}} \right\rangle ,\label{finitexy}
\end{aligned}
\end{equation}
and thus
\begin{equation}
\begin{aligned}
&[\langle n{\rm{(}}{\vec p}{\rm{)}}|{v_x}\left| {m{\rm{(}}{\vec p}{\rm{)}}}
\right\rangle \langle m{\rm{(}}{\vec p}{\rm{)}}|{v_y}\left|
{n{\rm{(}}{\vec p}{\rm{)}}} \right\rangle  + \langle n{\rm{(}}{\vec p}{\rm{)}}|{v_y}\left|
 {m{\rm{(}}{\vec p}{\rm{)}}} \right\rangle \\
 & \cdot \langle m{\rm{(}}{\vec p}{\rm{)}}|{v_x}\left|
 {n{\rm{(}}{\vec p}{\rm{)}}} \right\rangle ] = \frac{i}{\hbar }{e_{nm}}[\langle n{\rm{(}}{\vec p}{\rm{)}}|x\left| {m{\rm{(}}{\vec p}{\rm{)}}}
\right\rangle \\
&\cdot\langle m{\rm{(}}{\vec p}{\rm{)}}|{v_y}\left| {n{\rm{(}}{\vec p}{\rm{)}}}
 \right\rangle  - \langle n{\rm{(}}{\vec p}{\rm{)}}|{v_y}\left| {m{\rm{(}}{\vec p}{\rm{)}}}
 \right\rangle \langle m{\rm{(}}{\vec p}{\rm{)}}|x\left| {n{\rm{(}}{\vec p}{\rm{)}}} \right\rangle ].
\label{finitejinyibu}
\end{aligned}
\end{equation}
The factors ${e_{nm}}$ cancel each other, noticing $\sum\limits_m
{\left| {m{\rm{(}}{\vec p}{\rm{)}}} \right\rangle \langle
m{\rm{(}}{\vec p}{\rm{)}}| = \texttt{I}}  $,  we have
\begin{eqnarray}
{\sigma _1}(\omega ) = \frac{{{e^2}}}{{\omega V\hbar }}
\sum\limits_{{\vec p},m \ne n} {f_n({\vec p})} \langle
n{\rm{(}}{\vec p}{\rm{)}}|[{v_y},x]\left| {n{\rm{(}}{\vec p}{\rm{)}}}
\right\rangle  \equiv 0,\label{finite10}
\end{eqnarray}
since the commutator $[x,{v_y}]$ vanishes,  $\sigma _{\mu \nu
}^{(0)} \equiv {\sigma _2}$.  As for the second term
(\ref{finitesigma12}), simple algebra yields
\begin{equation}
\begin{aligned}
\sigma _{\mu \nu }^{(0)} = \frac{{{e^2}\hbar }}{V} \sum\limits_{{\vec p},m
\ne n} {\frac{{f_{mn}{\rm{Im}}[\langle
m{\rm{(}}{\vec p}{\rm{)}}|{\upsilon _\mu }\left| {n{\rm{(}}{\vec p}{\rm{)}}}
\right\rangle \langle n{\rm{(}}{\vec p}{\rm{)}}|{\upsilon _\nu }\left|
{m{\rm{(}}{\vec p}{\rm{)}}} \right\rangle ]}}{{e_{nm}^2}}}. \label{finite2}
\end{aligned}
\end{equation}
where $f_{mn}=[f_m({\vec p}) - f_n({\vec p})]$. Therefore the finite-temperature Hall conductance  for open system
is given by ${\sigma _{\mu \nu }} = \sigma _{\mu \nu }^{(0)} +
\hbar\Gamma \sigma _{\mu \nu }^{(1)}$. Following the same procedure,
we have
\begin{equation}
\begin{aligned}
\sigma _{\mu \nu }^{(1)} =& \frac{{{e^2}\hbar }}{{2V}}\sum\limits_{\vec p,m \ne n} {g(\theta )} {f_{mn}}{\rm{Re}}[\langle m(\vec p)|{\upsilon _\mu }\left| {n(\vec p)} \right\rangle \\
 &\cdot \langle n(\vec p)|{\upsilon _\nu }\left| {m(\vec p)} \right\rangle ]/e_{nm}^3 + \frac{{{e^2}}}{{4\omega V}}\sum\limits_{\vec p,m \ne n} {h(\theta )} {f_{mn}}\\
& \cdot \langle m(\vec p)|{\upsilon _\mu }\left| {n(\vec p)} \right\rangle \langle m(\vec p)|{\upsilon _\nu }\left| {n(\vec p)} \right\rangle /e_{nm}^2.
\label{finitettaoa}
\end{aligned}
\end{equation}
In the Heisenberg picture, the  velocity operator $(\mu = x,y,z)$ is
defined as
\begin{eqnarray}
{\upsilon _\mu } = \frac{i}{\hbar }[{H_S}({ p}),{r_\mu }] =
\frac{{{ p}}}{{{m^ * }}} + \frac{{\partial {d_\alpha }}}{{\partial
{p_\mu }}}{\sigma _\alpha }, \label{finitevelocity}
\end{eqnarray}
where the summations with respect to $\alpha$ automatically  assumed
in  the Einstein summation convention. With
Eq.~(\ref{finitevelocity})
\begin{eqnarray}
\langle m{\rm{(}}{\vec p}{\rm{)}}|{\upsilon _\mu }\left|
{n{\rm{(}}{\vec p}{\rm{)}}} \right\rangle  = \frac{{\partial {d_\alpha
}}}{{\partial {p_\mu }}}\langle m{\rm{(}}{\vec p}{\rm{)}}|{\sigma _\alpha
}\left| {n{\rm{(}}{\vec p}{\rm{)}}} \right\rangle ,\label{finitecelocitymu}
\end{eqnarray}
for $m \ne n$, it is easy to prove that in the  two-band model,
\begin{equation}
\begin{aligned}
{\rm{Im}}[\langle m{\rm{(}}{\vec p}{\rm{)}}|{\sigma _\alpha }
\left| {n{\rm{(}}{\vec p}{\rm{)}}} \right\rangle
\langle n{\rm{(}}{\vec p}{\rm{)}}|{\sigma _\beta }
\left| {m{\rm{(}}{\vec p}{\rm{)}}} \right\rangle ] =& m{\varepsilon _{\alpha \beta \gamma }}
\frac{{{d_\gamma }}}{d},\\
{\rm{Re}}[\langle m{\rm{(}}{\vec p}{\rm{)}}|{\sigma _\alpha } \left|
{n{\rm{(}}{\vec p}{\rm{)}}} \right\rangle \langle
n{\rm{(}}{\vec p}{\rm{)}}|{\sigma _\beta }\left| {m{\rm{(}}{\vec p}{\rm{)}}}
\right\rangle ] =&  - \frac{{{d_\alpha }{d_\beta }}}{{{d^2}}}
.\label{finiteshuang2}
\end{aligned}
\end{equation}
Collecting all together, we can obtain the  finite-temperature Hall
conductance Eq.~(\ref{totalsigma}) for open system.


\begin{thebibliography}{9}
\bibitem{note1}It is worth pointing out that  the Kubo's theory is developed for
the weakly non-equilibrium system, it can be applied to the open
system, but not for open quantum systems. For example,
Bloch-Gruneisen formula given by the Kubo formula is based on the
Boltzmann equation avaible for classical open system.
\bibitem{note2}The Mermin-Wagner(MW) theorem tells that
in one and two dimensions, continuous symmetries cannot be
spontaneously broken at finite temperature in systems with
sufficiently short-range interactions. If the MW theorem holds,  the
Hall conductivity is  absent in this system. However, the MW theorem
does not hold in our system, because the steady state here is not
the equilibrium state, i.e., $\rho=e^{-\beta H}$. So, the
Mermin-Wagner's theorem is not violated  by the example.
\end{thebibliography}
\end{document}